%% file: main.tex
\documentclass[11pt,epsfig,amsfonts]{article}
\usepackage{amsmath,amssymb,epsfig,eufrak,xcolor}
\usepackage{float, placeins}

\IfFileExists{kfont-charter.sty}{\usepackage{kfont-charter1}}{}

\renewenvironment{itemize}{
  \begin{list}{--}
    {\setlength{\parsep}{3pt}
      \setlength{\labelwidth}{24pt}
      \setlength{\itemsep}{1pt}
      \setlength{\topsep}{3pt}}}{\end{list}}

\usepackage{ifdraft}
\usepackage[numbers,sort&compress]{natbib}
\usepackage[pdfpagemode=UseNone]{hyperref}
\usepackage{nicefrac}
\renewcommand{\paragraph}[1]{\medskip\noindent{\bf{#1}}}

\usepackage{subcaption}

\restylefloat{table}
\usepackage[margin=1in]{geometry}

\usepackage{setspace}

\newcommand{\knote}[1]{}
\newcommand{\ktext}[1]{#1}
\newcommand{\ynote}[1]{}
\newcommand{\ytext}[1]{#1}
\newcommand{\xnote}[1]{}

\newcommand{\prob}{}
\def\prob(#1){\mbox{\bf P}({\it #1})}

\newcommand{\kpart}[1]{\part*{\uppercase{\Large #1}}}

\title{\bf Mostly-monocular responses and other visual functions in a
  multiscale network model of Macaque V1}

\author{Zhuo-Cheng Xiao\thanks{Departments of Mathematics and
    Neuroscience, New York University Shanghai.} \and Kevin
  K. Lin\thanks{Department of Mathematics, University of Arizona.} \and
  Lai-Sang Young\thanks{Corresponding author: {\bf lsy1@nyu.edu}.
    Courant Institute, New York University.}}

\begin{document}

\maketitle

\kpart{Summary}

Visual signals from the two eyes merge gradually as they pass through the primary visual cortex (V1).  Here we use a computational model of Macaque V1 to study the first stage of this integration \ktext{along the magnocellular pathway}, in layer 4C$\alpha$, aiming to infer neuroanatomical origins of binocular response.  \ktext{It is known that neurons in layer 4C$\alpha$ are predominantly monocular, \ytext{though} some do exhibit varying degrees of binocularity.  We find (1) the emergence of narrow binocular strips along borders of ocular dominance columns (ODC), a finding that aligns with experiments; (2) most consistent with data is when $10-30\%$ of interactions near ODC boundaries are cross-columnar; and (3) feedback from layer 6 is largely monocular.}  These results were obtained through systematic hypothesis testing using a multiscale model that is orders of magnitude faster than its biologically-detailed predecessors.  We propose that multiscale modeling can be an effective tool for bridging anatomy and function.


\newpage
\kpart{Introduction}
 
In primates, visual signals remain separate before they enter the visual cortex. The aim of this paper is to study the neural substrate that gives rise to the emergence of binocular responses. We do this via computational modeling, using a newly constructed multiscale model of the Macaque V1 cortex.  Our secondary aim is to call attention to multiscale modeling as a useful tool for hypothesis testing in investigations that involve large-scale neuronal simulations.

\medskip

Biologically realistic computational models can elucidate mechanisms and generate testable predictions~\cite{abbott2008theoretical, gerstner2012theory, sompolinsky2014computational, einevoll2019scientific, herz2006modeling,bassett2018nature, potjans2014cell}; they complement well modern \ktext{data-driven methods~\cite{yamins2016using,billeh2020systematic,eriksson2022combining,chen2022data,dura2024large,luo2025mapping}}.\knote{added dicarlo~\cite{yamins2016using}} \ktext{However, the complexity of biologically realistic} models poses formidable challenges that often render them infeasible or impractical~\cite{einevoll2019scientific,billeh2020systematic,markram2015reconstruction}.  Computational speed is critical in theoretical investigations when it is necessary to evaluate quickly large sets of hypotheses.  In \cite{xiao2024efficient} we developed a mathematical framework for multiscale simulations of cortical circuits.  The core idea was that by trading in certain network details, e.g., neuron-to-neuron variability in local populations, one could achieve substantial speedup.  We believe this modeling approach could be a very useful discovery tool in neuroscience, and demonstrate that via a concrete problem in this paper.

Our challenge here is to discover the mechanistic origins of binocular responses in the Macaque, whose visual system is similar to that of humans \cite{tootell2003neuroimaging,orban2004comparative}.  We present a new multiscale model of the Macaque visual cortex, specifically the input layer (4C$\alpha$) of V1 along the magnocellular pathway \cite{chatterjee2003parallel,nassi2009parallel}.  Our model employs the computational strategy in \cite{xiao2024efficient}, and follows the anatomical facts and physiological constraints used in the large-scale biologically-detailed computational models in \cite{chariker2016orientation,chariker2020contrast,chariker2022computational}.  Like \cite{chariker2016orientation,chariker2020contrast,chariker2022computational}, our model is capable of a broad range of basic V1 functions such as orientation selectivity and contrast response \cite{hubel1968receptive,ringach2002orientation,carandini2012area,priebe2012mechanisms}.  After establishing that the model performs reasonably in multiple visual tasks, we use it to infer anatomical structures and mechanisms in relation to the merging of signals from the two eyes.

It is well established that signals from the two eyes remain separate before entering V1, and that cells that receive signals from the left and right eyes are gathered in distinct regions, or stripes, called ocular dominance columns (ODCs) \cite{Tootell1988a,adams2009ocular}. Experiments have also shown that neurons in the input layers of V1 are largely -- though not completely -- monocular \cite{hubel1968receptive,hawken1984contrast,Tootell1988a,katz1989local}. We posit, as others have, that the origin of binocularity has to do with interaction across ODC borders.  But many questions remain: Does cross-ODC interaction already happen in the input layers of V1, or does it start later? What is the extent and nature of that interaction? How do ODCs impact the microcircuitry that underlies binocular integration, and how does feedback from deeper layers contribute to the computation \cite{katz1989local,angelucci2006contribution,dougherty2019binocular,maier2022binocular}?

This emergence of binocular responses in neurons was not considered in \cite{chariker2016orientation,chariker2020contrast,chariker2022computational} (which modeled only one eye), nor have these issues been addressed in any previous theoretical study that we know of.  We will show that our multiscale model is an effective tool for addressing these and other questions.

\kpart{Results}

First we present a multiscale model of Layer 4C$\alpha$ of V1 {\em without ODCs}, and verify that it captures V1's most salient features.  Then we \ktext{add ODCs to the model, using it to} investigate the nature and extent of cross-ODC interactions, examining through simulations the consequences of different connectivity patterns.  Finally, we combine these predictions with existing data to propose a likely picture.

\section*{A computational model exhibiting orientation selectivity,
  contrast response, and time frequency selectivity}

Layer 4C$\alpha$ (L4) is the input layer of the magnocellular pathway to V1 \cite{chatterjee2003parallel}.  It receives feedforward input from LGN and feedback from Layer 6 (L6) \cite{wiser1996contributions,sincich2005circuitry}.  As depicted schematically in Fig.~1A, one often divides V1 into {\it hypercolumns} (HCs) consisting of cells that share roughly the same receptive fields; each HC is further subdivided into {\it orientation domains}, collections of cells with like orientation preferences~\cite{hubel1962receptive,blasdel1992orientation,obermayer1993geometry,lund2003anatomical}. In Fig.~1A, HCs in L4 are depicted as squares, and orientation domains are marked by triple bars indicating \ktext{their} preferred orientations.  Connections between neurons within L4 are short-ranged, and local network architectures are similar independent of spatial location \cite{callaway1998local}.

\paragraph{\ktext{Preliminary} model without ODCs.} 
In real V1, signals from the two eyes collect in bands called ocular dominance columns (ODC). We first build a model {\it without} ODCs (Figs.~1--3). The reason for omitting ODCs to begin with is that their modeling involves additional issues, the very issues we wish to address. We use this \ktext{preliminary} model to calibrate physiological parameters unrelated to dominance columns; ODCs will be inserted later on.

\begin{figure}[tp!]
  \begin{center}
    \ifdraft{}{\includegraphics*[width=\textwidth]{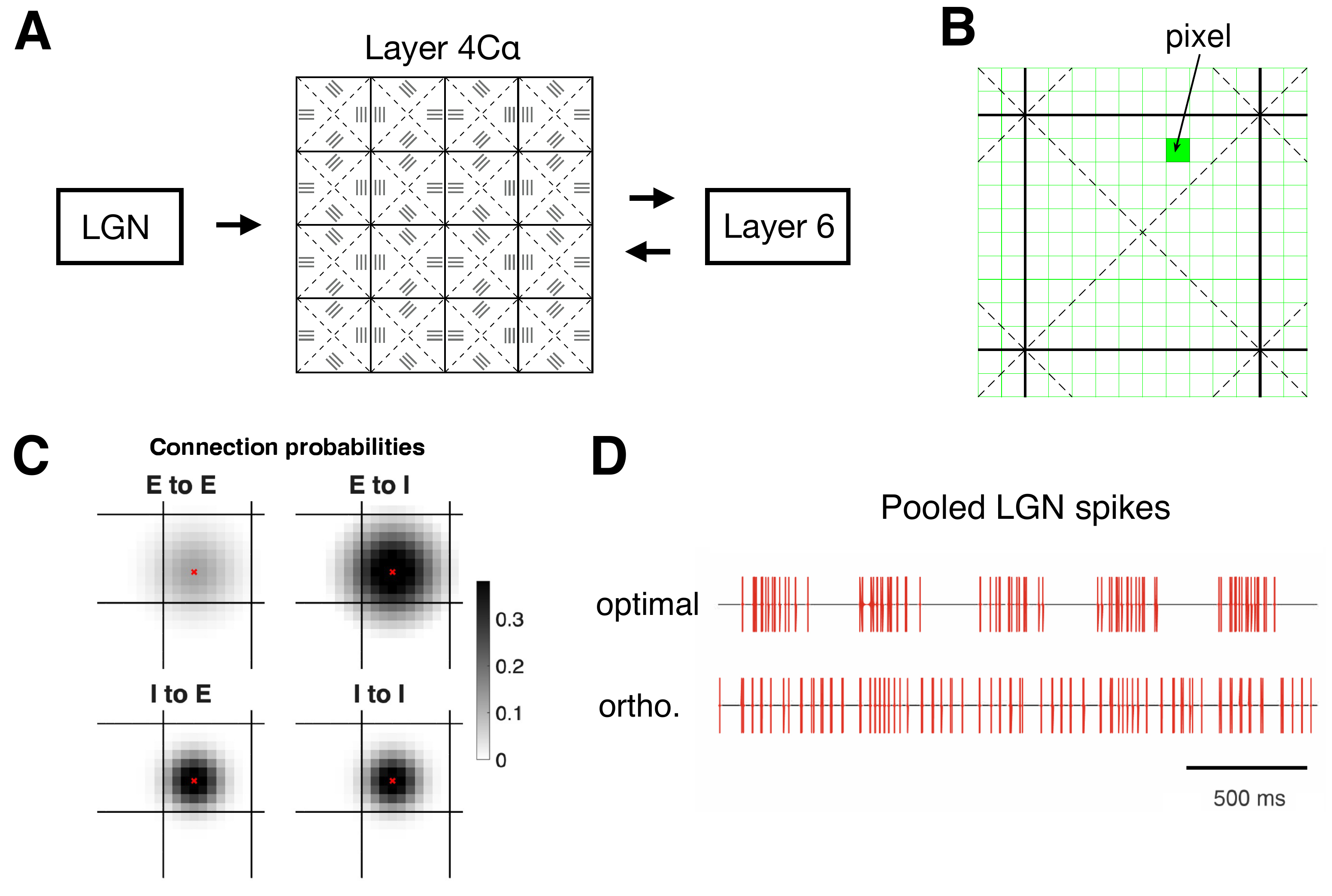}}
    \caption{{\bf Model layout.}~~\textbf{A.} Schematic of Layer 4C$\alpha$ (L4) with feedforward input from LGN and feedback from Layer 6.  Note as explained in the text, we defer treatment of ODCs until later (see Fig.~4). L4 is subdivided in hypercolumns (squares) consisting of neurons with similar receptive fields, and further subdivided into orientation domains; \ktext{triple bars indicate the orientation of LGN afferents for cells in that orientation domain.} \textbf{B.} Each hypercolumn of L4 is subdivided into $10 \times 10$ local populations of neurons; these are the basic units in the coarse-grained component of the multiscale model.  \textbf{C.} Summary of connection probabilities between local populations; for simplicity, we have lumped simple and complex cells together as E (excitatory) cells.  \textbf{D.} Spike trains of 4 LGN afferents to a simple V1 cell in response to a $10$ Hz grating pooled together.  When the grating is aligned with the ON/OFF rows of LGN (top), orthogonal to the rows of LGN (bottom).}
    \label{Fig1: Intro}
  \end{center}
\end{figure}

This model follows closely the underlying neuroanatomy in the biologically-detailed models \cite{chariker2016orientation,chariker2020contrast,chariker2022computational}, but computations are performed very differently. Instead of simulating directly the \ktext{$\sim10^5$} neurons in the model, we gain speed (by factors of up to $\mathcal O(10^3)$) using the multiscale methods developed in \cite{xiao2024efficient}. \ktext{We refer the reader to the SI and to \cite{chariker2016orientation} for the relevant neurobiology.}  Below we summarize the main steps in the new computational scheme.

\smallskip
{\em Multiscale simulations of L4 dynamics.}  We divide L4 into disjoint groups of nearby neurons called ``local populations".  Neurons within a local population interact with one another and respond to the inputs they receive from ``outside'' (meaning outside of the local population); we refer to this as {\it local dynamics}.  It was shown in \cite{xiao2024efficient} that local responses can be \ktext{modeled} efficiently thanks to the generic architecture of local circuits.  Complementary to local dynamics are {\it coarse-grained (CG) dynamics}, which describe interactions among local populations. CG dynamics take place on a phase space that is much reduced in dimension and are therefore \ktext{far less costly} to simulate.

In more detail, we divide each HC in our model into $10\times10$ local populations (Fig.~1B). Within each local population we distinguish 3 cell types: simple and complex E-cells, and I-cells. Interaction between local populations is deduced from connection probabilities in real cortex (see references in \cite{chariker2016orientation}) and are summarized in Fig.~1C.  The size a local population is chosen so that neurons of each type within it can be viewed as behaving similarly.  In this way, the \ktext{steady-state} response of our model to a visual stimulus can be summarized by the collection of triplets $$\{(f^S_p, f^C_p, f^I_p), \ p = \mbox{local population} \}\ ,$$ where $f^S_p$, $f^C_p$, and $f^I_p$ represent mean firing rates of simple, complex, and inhibitory cells in the $p$th local population.  Mean E-firing rate $f^E_p$ is given by $f^E_p = 0.7 f^S_p + 0.3 f^C_p$~, reflecting the proportions of simple and complex cells in each local population.  Updating procedures and the interplay between CG and local dynamics are briefly summarized in {\bf Methods}, with additional details in SI and \cite{xiao2024efficient}.

\smallskip
{\em External inputs to L4.}  Following Hubel and Wiesel~\cite{hubel1968receptive}, we wire the simple cells in V1 to spatially aligned rows of ON and OFF LGN cells \ktext{(see SI)}.  \ktext{As LGN input to L4 is assumed to be purely feedforward, LGN dynamics need not be computed in real time, and it is more efficient to precompute LGN responses to various stimuli and feed them as inputs to the L4 model above.  Specifically,} given a visual stimulus, we compute the pooled LGN input, i.e., the superposition of spike trains from all LGN cells presynaptic to a V1 cell.  Responses of LGN cells to typical V1 cells for a collection of stimuli that include drifting gratings with a range of orientations and spatial/temporal frequencies are precomputed and stored; see Fig.~1D for two examples of pooled LGN spike trains.  L6 inputs to L4, which are assumed to reflect approximately activity levels of L4 at corresponding locations, can also be precomputed.  See {\bf Methods} and SI for details.

\medskip
In the rest of this paper, by E and I-firing rates at specific locations in L4, we mean the {\it mean firing rates $f^E$ and $f^I$ of the local population} at that location.  We now present a sample of results to illustrate the model's capabilities.

\paragraph{Orientation selectivity (OS).} 
OS is one of the most basic functions of V1 cells, yet the \ktext{underlying mechanisms that lead to} observed orientation preferences, are subtle. First, OS is driven not by firing rates but by LGN firing {\it patterns} \cite{ferster1996orientation}: the spatial alignment of a V1 cell's LGN afferents relative to the orientation of the grating presented produces LGN spike trains with different patterns (see Fig.~1D), and these patterns elicit vastly different responses from the V1 cell.  Second, \ktext{\ytext{with} so few  orientation domains} \ytext{in each HC (and correspondingly few distinct LGN-to-V1 alignments)}, V1 cells are able to discern a nearly continuous range of angles around the clock, a capability the model must be able to replicate.

Fig.~2A shows example tuning curves for two local populations: the first (``Population 1'') is located near the center of a vertical-preferring domain while the second (``Population 2'') is on the border between the $0^\circ$ and the $45^\circ$-preferring domains.  As with individual neurons, for each local population we record its E-firing rate in response to a drifting grating as the angle of the grating is varied in small steps. Both local populations show strong OS, with Population 1 (preferring a cardinal direction) more tuned. Tuning curves for other local populations are similar.

The rest of Fig.~2 shows {\it activity maps} in response to various drifting gratings.  In an activity map, the color of each pixel indicates the firing rate (E or I) of the local population, enabling one to assess at a glance activity levels across the entire model cortex.  We have found these maps to be useful diagnostic tools and will use them repeatedly in \ktext{this paper}.

To verify that the wiring between LGN and cortex (following the hypothesis of Hubel and Wiesel) produces the intended orientation preferences, readers should compare the ``hot spots'' in Fig.~2B, i.e., regions with brighter colors and higher firing rates, to the intended orientation domains shown in Fig.~1A.  For example, when the $0^\circ$ grating is presented, local populations in the domain marked with vertical bars fire the most vigorously, and populations in the domains preferring horizontal are the least active, and likewise when\ktext{, e.g.,} $45^\circ$ gratings are presented.  To help picture how activity level changes from the map at $0^\circ$ to the one at $45^\circ$, we provide 3 snapshots in between, at $15^\circ$, $22.5^\circ$, and $30^\circ$, confirming that with only 4 sets of LGN alignments, the model is able to produce a wide range of activity patterns.

Fig.~2C shows the corresponding activity maps for I-cells. We confirm that as in real cortex, E and I-spiking in the model covary, i.e., I-firing rates are high when E-firing rates are high and {\it vice versa}, I-firing rates are $\sim3$ times higher, and I-cells are more broadly tuned \cite{swadlow1988efferent,cardin2007stimulus}.

\begin{figure}[tp!]
  \begin{center}
    \ifdraft{}{\includegraphics*[width=\textwidth]{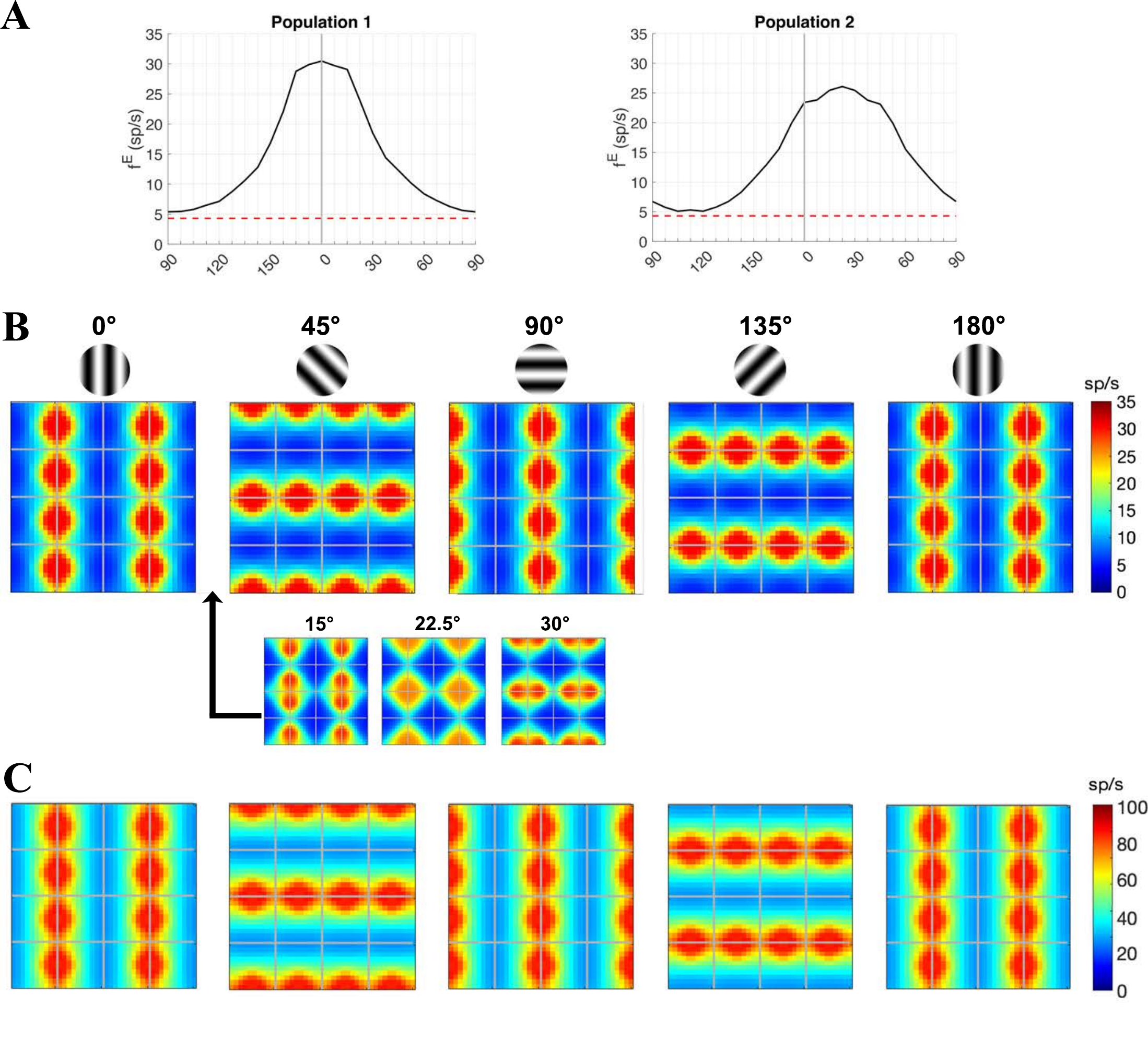}}
    \caption{{\bf Orientation selectivity: tuning curves and activity maps.}~~ \textbf{A.} Example tuning curves for two local populations in the model described in Fig.~1.  Population 1 is located near the center of a vertical-preferring domain; Population 2 lies on the border of the domains preferring $0^\circ$ and $45^\circ$.  Horizontal dashed lines indicate background firing rate.  \textbf{B.} Activity maps of E-cells, showing outputs of the coarse-grained model in response to drifting gratings with various angles (temporal frequency = 10Hz, spatial frequency = 2.5 \ktext{cycles/deg}, contrast = 64\%); the E-firing rate of each local population is indicated by color.  Activity maps for grating angles between $15-30^\circ$ are shown in the inset. \textbf{C.} Corresponding activity maps for I-cells.}
    \label{Fig2:Model outputs}
  \end{center}
\end{figure}

\paragraph{Contrast response.}
A salient feature of magno-driven V1 cells is their strong sensitivity to contrast.  When presented with gratings with increasing contrast, firing rates of these V1 cells rise steeply, saturating at fairly low contrast \cite{albrecht1982striate,henry2013functional}.  As with OS, this property is also more subtle than meets the eye: It has been established that LGN currents comprise only a small fraction of total excitatory current received by V1 cells \cite{douglas1995recurrent}, and that their outsize influence on their postsynaptic V1 cells is a consequence \ktext{of E-I balance within cortex (see the analysis in \cite{chariker2020contrast})}.  We show here that the model correctly reproduces these features.

The two curves in Fig.~3A show firing rates as functions of contrast for the same two local populations as in Fig.~2A. For Population 1, the curve rises steeply, passing {\small$\frac12$}-max at $\sim10\%$ contrast and saturating quickly.  The curve for Population 2 rises more slowly, reaching {\small$\frac12$}-max between 10 and $25\%$ contrast; this likely has to do with the fact that Population 2 receives mixed LGN input with two different spatial alignments. These two curves represent the two extremes: contrast responses curves for local populations at intermediate distances to the boundary of their orientation domains lie between these two.
  
\begin{figure}[tp!]
  \begin{center}
    \ifdraft{}{\includegraphics*[width=\textwidth]{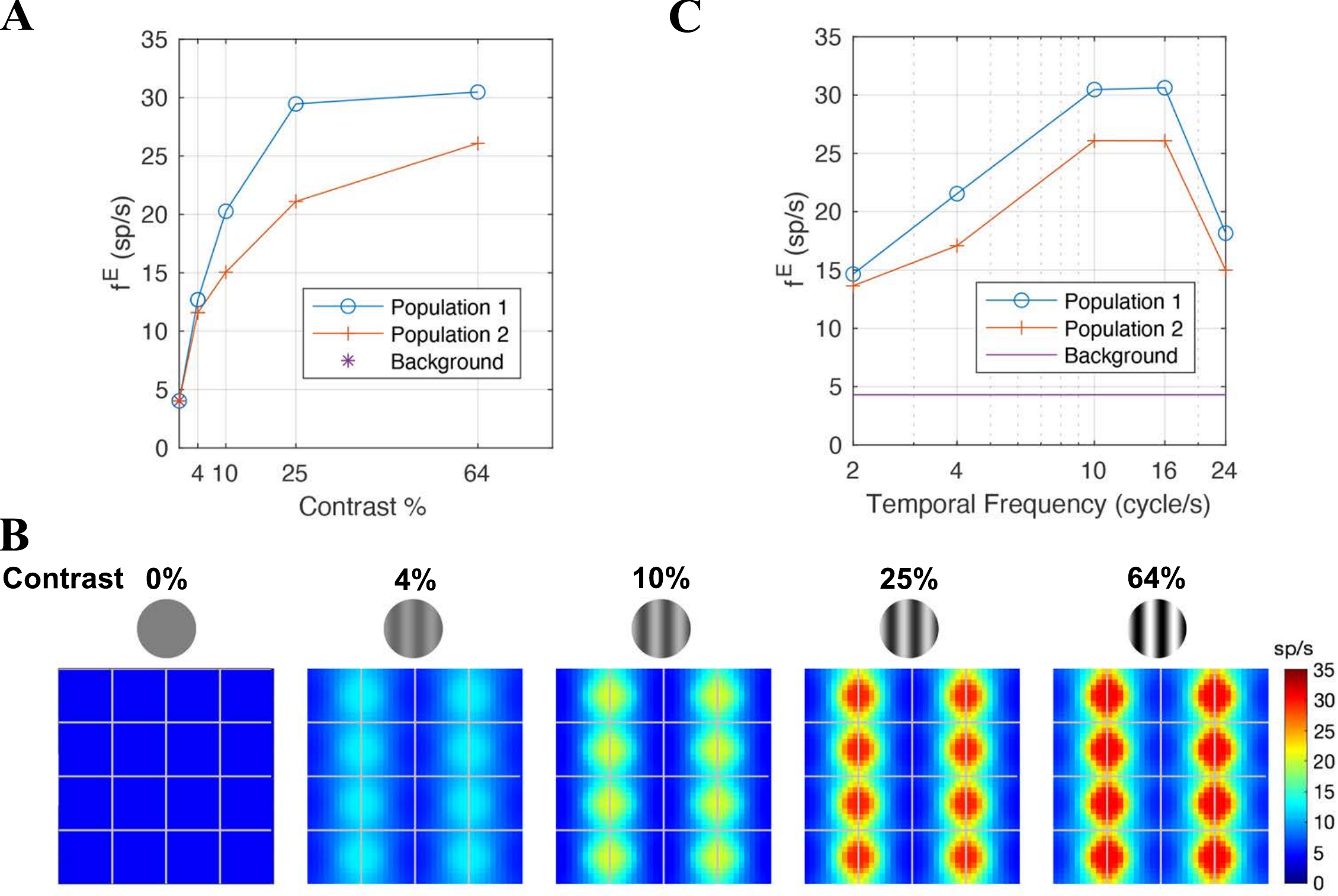}}
    \caption{{\bf Two other V1 properties.}~~ \textbf{A.} Contrast response: firing rates as function of contrast are plotted for the same two local populations as in Fig.~2A ({\Large\textcolor{blue}{$\circ$}}: Population 1, \textcolor{red}{$+$}: Population 2); ``$0\%$ contrast'' = background.  \textbf{B.} Activity maps in response to the vertical grating at several different contrasts.  \textbf{C}. Firing rates as functions of the grating's temporal frequency ({\Large\textcolor{blue}{$\circ$}}: Population 1, \textcolor{red}{$+$}: Population 2).}
    \label{Fig3:Model outputs}
  \end{center}
\end{figure}

Fig.~3B shows activity maps corresponding to several different contrasts for a vertical grating.  Here we see that the regions of elevated activity remain roughly unchanged as contrast varies, demonstrating the principle of approximate contrast invariance \cite{albrecht1982striate,carandini2004contrast,henry2013functional}.

\paragraph{Temporal frequency (TF) selectivity.}  
This property is chosen to bring home the point that our model is capable of capturing temporal properties of visual inputs.  Through the use of recorded spike trains, LGN responses to drifting gratings at different temporal frequencies are conveyed to their postsynaptic V1 cells.  Results of the model's responses to gratings at four TFs are shown in Fig.~3C. They are in good agreement with data, which show that L4 E-cells typically prefer TF between 8 and 16 Hz \cite{hawken1996temporal}.

\bigskip
Having established that the model reproduces satisfactorily 
a number of basic V1 properties, we now extend it to include ocular dominance columns
and use the resulting model to carry out a theoretical study of 
the mechanistic origin of binocularity.

\section*{Where and how do visual signals from the two eyes merge?}
For the Macaque, it is well established that visual signals from the two eyes go to separate layers of the LGN \cite{connolly1984representation,callaway2005structure}, which project to well-segregated, alternating eye-specific bands in the input layers of V1. These bands form the anatomical basis of the physiologically observed ocular dominance columns \cite{horton1998monocular}.  It has also been established (through C-2-deoxyglucose labeling) that the density and definition of the boundaries of ODCs vary with laminae, the columns being the darkest with most sharply defined boundaries in the input layers, i.e. layer 4C (and Layer 6), lighter with less well defined boundaries in output layers, i.e., layers 2, 3 and 4B (as well as 5 and 4A) \cite{lund2003anatomical}. These findings correspond well with measurements of the binocularity of individual neurons \cite{hubel1968receptive,hawken1984contrast,Tootell1988a,katz1989local}:
in the input layers, neurons have been observed to be largely though not exclusively monocular, whereas neurons in the output layers are very binocular \cite{hubel1968receptive,hawken1984contrast,cox2019temporal, dougherty2019binocular}. 

The merging of signals likely occurs both through interlaminar and intralaminar connections.  Below we investigate to what degree this merging occurs within Layer 4C$\alpha$, and how. That leads naturally to questions of microcircuitry, to the extent to which neurons in an ODC interact with those in the ``opposite column'' in this input layer.  On these questions, we have drawn inspiration from \cite{katz1989local}, which reports on an experimental study of neurons in Layer 4C$\beta$.  The authors showed, with examples, that dendrites of cells close to ODC borders tend to remain in the home column; axonal collaterals could cross into adjacent columns, but their lateral spread has remained largely in the home column.  (See also \cite{katz1989local}.)  We are not aware of results for 4C$\alpha$, and \ktext{will use our model} to investigate the anatomical structures in relation to ODC in this layer, which is thought to be less monocular than 4C$\beta$ though still largely monocular~\cite{hawken1984contrast, Tootell1988a}.

\paragraph{\ktext{Extended model} with ocular dominance columns.} 
To carry out this study, we must first extend the model presented above to include ODCs.  Fig.~4A shows the layout of the new model.  Each of the 4 rows of HC in the previous model is replaced by two rows labeled “L” and “R”, indicating the eye \ktext{from} which input signals originate.  ``L'' and ``R'' rows of HC alternate as shown; adjacent rows have displaced but overlapping receptive fields.

Following \cite{blasdel1992orientation,obermayer1993geometry}, we assume that pinwheel centers lie roughly midway between two ODC boundaries. We also assume that some but not all orientation domains straddle two columns.  The assignment of the domains preferring cardinal directions to lie in the interior of an ODC and the diagonal-preferring ones to be on ODC borders was an arbitrary choice on our part --- we were unable to locate anatomical guidance.  In our analysis below, it is the distance to an ODC border that matters, not the actual angle preferred.

L4 neurons receive synaptic inputs from three sources: they interact with other neurons within L4, receive feedforward input from LGN, and feedback from L6.  \ktext{LGN input is modeled as before, except neurons in ODCs labeled ``L'' and ``R'' receive input from two separate layers of LGN.  The modeling of projections from L6 is unchanged: these projections are assumed to be oblivious to ODC borders, based on the experimental observation that inter-laminar axonal projections tend not to respect ODC boundaries \cite{wiser1997ocular}.  The most significant changes are to connection probabilities within L4, which we now describe.}

\paragraph{Crossing rules.}
          {\it A priori,} ocular columns may impact anatomical structure in the following ways: neurons may interact across ODC boundaries unobstructed, or axonal arbors and/or dendritic trees may preferentially remain in the home column (\ktext{as has been observed in Layer 4C$\beta$} \cite{katz1989local}). To these, we add a third possibility, namely that density of connections near ODC borders could be lower due to aversion (in development or through pruning) to cross-column interaction.

\ytext{To formalize these scenarios, we fix 3 nonnegative numbers summing to 1:
\begin{align*}
  \prob(Crs) &~=~ \mbox{probability of ``crossing,''}\\[1ex]
  \prob(Rfl) &~=~ \mbox{probability of being ``reflected,'' and}\\[1ex]
  \prob(Abs) &~=~ \mbox{probability of being ``absent.''}
\end{align*}
Let $p$ and $q$ be local populations and $P$ and $Q$ two cell types, i.e., $P,Q\in\{E, I\}$.  Let $C_{Q\leftarrow P;q\leftarrow p}$ denote the average number of cells of type $P$ in population $p$ that project to each cell of type $Q$ in population $q$, following the preliminary model depicted in Fig.~1C. \ktext{Note these numbers encode network structure in our CG model (see SI).}  For each choice of $\{\prob(Crs), \prob(Rfl), \prob(Abs)\}$, we modify the collection of numbers $\{C_{Q\leftarrow P;q\leftarrow p}\}$ to $\{\hat C_{Q\leftarrow P;q\leftarrow p}\}$ as follows: if $p,q$ lie in opposite ODCs with $C_{Q\leftarrow P;q\leftarrow p} > 0$, we set
\begin{align*}
  \hat C_{Q\leftarrow P;q\leftarrow p} &~=~ \prob(Crs) \times C_{Q\leftarrow P;q\leftarrow p}~~\mbox{and}\\[1ex]
  \hat C_{Q\leftarrow P;q’\leftarrow p} &~=~ C_{Q\leftarrow P;q’\leftarrow p} + \prob(Ref) \times C_{Q\leftarrow P;q\leftarrow p}~,
\end{align*}
where $q’$ is the mirror image of $q$ in the ODC of $p$ (see Fig.~4B). That is, of the $C_{Q\leftarrow P;q\leftarrow p}$ connections from population $p$ to each neuron in population $q$, a fraction of $\prob(Crs)$ is untouched, a fraction $\prob(Rfl)$ is reflected back to $p$'s home column, and the rest (a fraction $\prob(Abs)$) are deleted.  The same rule holds for all pre- and postsynaptic cell types.  Connections between local populations in the same ODC are untouched.  This algorithm produces a new network for each choice of $\{\prob(Crs),\prob(Rfl),\prob(Abs)\}$, but as we will show, not all resulting models will have reasonable properties.  Our task is to identify those choices of $\{\prob(Crs),\prob(Rfl),\prob(Abs)\}$ that are compatible with experimental data.}

\bigskip
We summarize our model \ktext{findings} in the 5 points below, along with mechanistic \ktext{analyses}.

\paragraph{Finding 1: Density of connections near ODC borders is the same as elsewhere, i.e. $\prob(Abs)$ is close to $0$.}  We start by scanning the model's responses to monocular stimulation as crossing rules are varied. From the symmetries in HC organization, it suffices to show these responses from 4 HCs. Fig.~4C shows activity maps for the 4 HCs in the red square in Fig.~4A \ktext{as they respond} to a vertical ($0^\circ$)-drifting grating presented to only one eye, the eye that projects to the top two HCs.  The $x$- and $y$-axes are self-explanatory, e.g., $\prob(Crs)=0.5$ and $\prob(Rfl)/(\prob(Rfl) + \prob(Abs))=0.25$ means that a potential inter-ODC connection is present with probability $0.5$, is reflected with probability $0.125$, and is deleted otherwise.

\begin{figure}[tp!]
  \begin{center}
    \ifdraft{}{\includegraphics*[width=\textwidth]{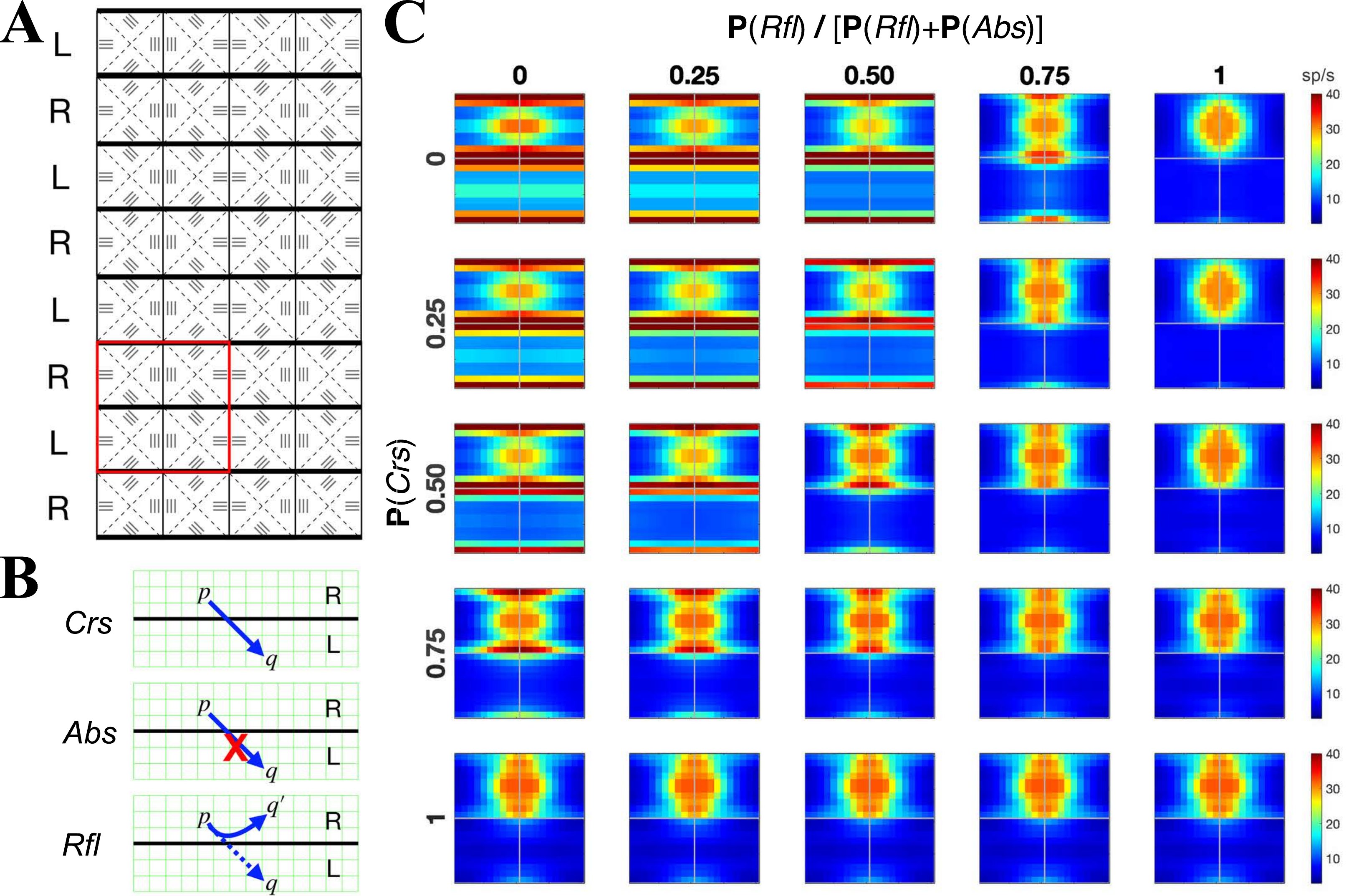}}
  \end{center}
  \caption{{\bf Two-parameter family of computational models with ocular dominance columns (ODCs).}~~ \textbf{A.} Schematic of a model with 8$\times$4 HCs; alternate rows correspond to ODCs receiving input from the left (L) and right (R) eyes.  \textbf{B.} The three crossing rules, depicting how a connection from \ktext{a cell in population $p$ to a cell in population $q$  that potentially crosses ODC boundaries may be modified (if it is present according to previously defined connection probabilities).}  \textbf{C.} Activity maps of E-firing rates for the 2$\times$2 HCs in the red box of \textbf{A} under monocular stimulation.  The visual stimulus is a vertical ($0^\circ$ degree) drifting grating presented to the eye corresponding to the upper two squares; the other eye is occluded.  The probabilities in the $x$ and $y$-axes together describe the full 2-parameter family of crossing rules; see text for detail.}

  \label{Fig4:Cross-and-reflect}
\end{figure}

Model responses in the upper left quadrant of Fig.~4C are strikingly unrealistic: local populations on both sides of the ODC border fire at very high rates, up to 60-70 sp/s (cut off in the figure), up from \ytext{a maximum of} 30-35 spikes/s during binocular stimulation.  We note further that these high-firing local populations are in orientation domains designed to prefer 30-60$^\circ$ and 120-150$^\circ$, not the orientation of the grating, and nearly half of them are in the column of the unstimulated eye.  The same defects, to a lesser degree, are found in all panels where $\prob(Abs)$ is significantly above $0$\ktext{.  The} rightmost column, corresponding to $\prob(Abs) =0$, shows the most realistic activity maps, meaning maps resembling those in Fig.~2.

A similar study was carried out for the $45^\circ$ grating, with similar results; see SI.

\smallskip
{\it Analysis.} These results may appear counter-intuitive: for local populations near the ODC border, when $\prob(Abs)$ is large, a substantial fraction of connections between L4 neurons is missing, resulting in the loss of a significant portion of current from within L4.  It is tempting to jump to the conclusion that this should result in lower, not higher, firing rates.  {Why, then, are firing rates higher?  The reason is that had all cross-ODC connections been present, the sum of E and I-currents from within L4 received by local populations would in fact be {\it \ktext{net} negative} \knote{we don't discuss ness in si}\ktext{(see SI)}, and missing a fraction of a net-negative current causes firing rates to go up.  At $\prob(Crs)=0$ and $\prob(Abs)=1$, \ktext{i.e.,} with about half of the connections missing, \ktext{these results show that} lateral interaction alone causes firing rates along the ODC border to be elevated\ktext{, and} recurrent interaction with L6 further amplifies the elevated E-firing.}

A remark on the sign of net cortical currents: while cells in L4 receive fluctuating, net-negative E and I currents from within L4, feedforward input from LGN, feedback from L6, and ambient drive are all excitatory-only.  The total current from all sources combined is on average net-positive, and spiking activity in L4 is driven by this net-positive current.  That currents from within L4 are fluctuating and on average negative has been seen in other models (see, e.g., \cite{chariker2022computational}).

\bigskip
From here on we assume that $\prob(Abs)=0$, so $\prob(Crs) + \prob(Rfl) =1$.

\paragraph{Finding 2: Responses to monocular and binocular stimulations
vary with distance to ODC borders.}  To examine more closely the behavior of local populations, we plot their tuning curves, a sample of which is shown in Fig.~5. We consider here 6 local populations at several different values of $\prob(Crs)$. Three sets of tuning curves describing the populations' responses to binocular stimulation (black), monocular stimulation (red), and monocular stimulation of the ``other eye'' (blue) are shown, ``other eye'' meaning the eye corresponding to the ODC that is not the local populations's home column.

The local population at location {\bf a} (``{\bf Pop.~a}'') in Fig.~5, is located  far from an ODC border. It has the properties that (i) when the other eye alone is stimulated, its mean firing rates are only marginally above background levels, and (ii) when its ``home column'' is stimulated, it produces nearly identical responses to monocular and binocular stimulations.  This is generally what is expected of a monocular neuron.

Deviation from ideal monocular behavior increases as we move towards an ODC border.  \ktext{These} ``defects'' are mild at locations {\bf b} and {\bf f}, becoming more pronounced at {\bf c}, {\bf d}, and {\bf e}.

\begin{figure}[tp!]
  \begin{center}
    \ifdraft{}{\includegraphics*[width=0.7\textwidth]{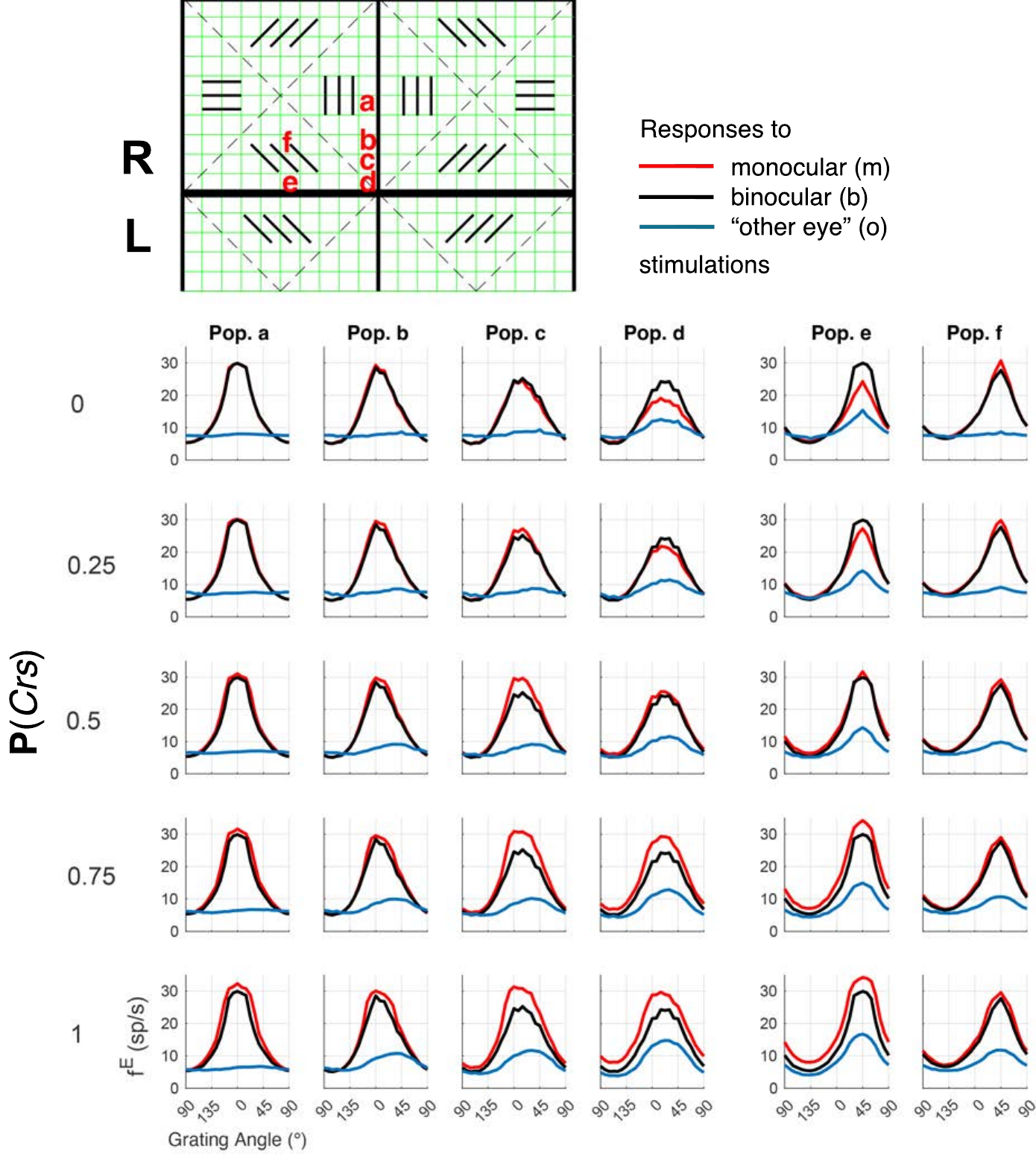}}
  \end{center}
  \caption{{\bf Example tuning curves under monocular and binocular stimulations.}~~Three tuning curves are shown for each of 6 local populations ({\bf Pops.~a-f}) for various values of $\prob(Crs)$.  Locations of the local populations are indicated in the top panel, which shows the same 4 HCs in the red square in Fig.~4A.  The 3 tuning curves correspond to monocular stimulation of the home column of the pixel (red), binocular stimulation (black), and monocular stimulation of the opposite column (blue).}
 \label{Fig5:Tuning-curves}
\end{figure}

\smallskip
{\it Analysis.} {We study systematically $b$ (binocular) and $m$ (monocular) responses in {\bf Pop.~e} \ktext{(see Fig.~5)}.

\ktext{First} observe that $b$ is independent of $\prob(Crs)$: the fraction of E/I-inputs from the opposite ODC vary with $\prob(Crs)$ but total E/I-inputs from within L4 remain unchanged.  \ktext{Now to see why $m$ increases with $\prob(Crs)$, note under monocular stimulation, input currents coming from within L4 decrease in magnitude as $\prob(Crs)$ increases due to the increasing fraction of connections from the opposite ODC (where firing rates are lower).  But these currents are net-negative (Finding~1),} so lateral interaction favors increasing firing rates as $\prob(Crs)$ increases.  As for L6, input from L6 is indexed to the average activity in {\bf Pop.~e} and its neighbors.  Firing rates of neighbors in the opposite ODC are low, so {\bf Pop.~e} and its neighbors in the home column (which behave like {\bf Pop.~e}) dominate.  Together this causes $m$ to increase with $\prob(Crs)$.

Finally, we compare $m$ and $b$. At $\prob(Crs)=0$, L4 dynamics in the home ODC are identical in the two cases, but feedback from L6 tips the scale to produce $b>m$ as it reflects the driven dynamics in the opposite ODC when binocularly stimulated.  At $\prob(Crs)=1$, the current in cross-column interaction has larger magnitude when binocularly stimulated, leading to $b<m$.}

\paragraph{Finding 3: Emergence of a narrow binocular strip.} We define the {\it binocular index} (BI) 
of a local population to be\knote{changing $b,m,o$ to $\bar{b}$ etc}
$$
BI = \frac{\bar{o}}{\bar{o}+\bar{m}}~,
$$ where $\bar{m}$ is the mean firing rate above background -- maximized over gratings of all orientations -- when the home column of the population is monocularly stimulated, and $\bar{o}$ is \ktext{the} firing rate in response to the same grating when the other eye is stimulated.
(Our definition of BI is equal to $\frac12(1- BI_{HW})$ where $BI_{HW}$ is the definition given in \cite{hubel1968receptive}.)

BIs for all local populations of a single hypercolumn are plotted in Fig.~6. A striking feature is the existence of binocular strips: populations along the borders of the ODC have significantly higher BI than elsewhere in the hypercolumn, {and this property is valid across all $\prob(Crs)$.  In {\bf Pop.~e}, for example, observe from Fig.~5 that $\bar{o}\gtrsim\frac13~\bar{m}$ for all $\prob(Crs)$, consistent with $BI\gtrsim\nicefrac14$ as shown in Fig.~6; and $\bar{o}\approx\frac12~\bar{m}$ at $\prob(Crs)=0$ or $1$, consistent with $BI\approx\nicefrac13$.

The existence of binocular strips in our model is} consistent with experimental findings\ktext{.  Indeed,} a similar phenomenon was noted in \cite{levay1980development,levay1988ocular} and studied carefully in \cite{horton1998monocular}, which reported the existence of binocular strips $50$-$70 \mu m$ wide in monkey L4\ktext{; for comparison,} each local population in our model occupies a cortical space of $50 \mu m\times \ 50 \mu m$.  We remark that no attempt was made in our model design to create these binocular strips.  This is an emergent phenomenon --- it arises not as a direct consequence of local rules but as a result of interactions among populations of neurons.

\begin{figure}[tp!]
  \begin{center}
    \ifdraft{}{\includegraphics*[width=\textwidth]{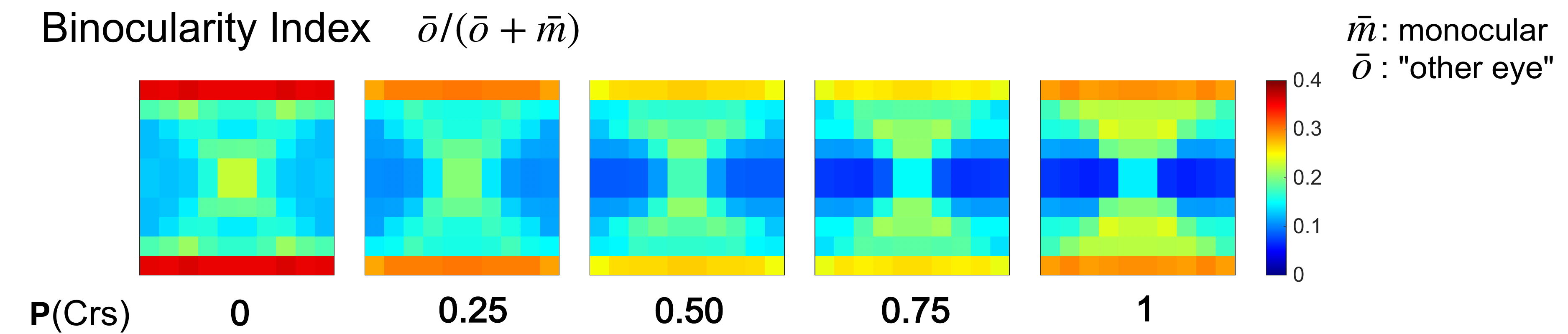}}
    \caption{{\bf Emergence of binocular strips.}  Binocular indices are depicted in color code for one hypercolumn, under a range of crossing rules.  \ktext{Color bar shows BI as defined in the main text: BI$=0$ means the local population is purely monocular; BI$=\nicefrac12$ means it is fully binocular.  Observe the emergence of binocular strips, as well as the fact that cells in orientation domains away from ODC borders (here horizonal and vertical-preferring) are the most monocular.}}
     \label{Fig6:Bino-index}
  \end{center}
\end{figure}

\smallskip 
{\it Analysis.} {here are two ways for information from the other eye to pass to the home ODC: via connections that cross ODC borders, or via L6. When the other eye alone is stimulated, cross-ODC connections (when present) carry a net-negative current to the home column.  Our simulations show that this negative current is outweighed -- by a substantial margin -- by the positive current from L6 in response to the elevated spiking in the opposite ODC.  Recall that L6 projections are oblivious to ODC borders, and the footprint of L6 projections originating from a border population in the opposite column extends roughly one row into the home column. These characteristics are consistent with both the high BI in the row adjancet to the ODC border and the abrupt drop in BI away from the border.

}

\paragraph{Finding 4: Binocular modulation suggests a majority -- but not all -- of potentially cross-ODC connections are reflected.}  It is well established experimentally that the responses of neurons to binocular and monocular stimulations do not differ substantially \cite{katz1989local,cox2019temporal}.  In Finding~2, we proposed that
the differential of these responses vary with $\prob(Crs)$.  To investigate this systematically, we define (following \cite{dougherty2019binocular,mitchell2022stimulating}) the {\it binocular modulation} (BM) of a local population to be
$$ BM = \frac{\bar{b}-\bar{m}}{\bar{b}+\bar{m}}\ ,
$$ \ktext{where $\bar{m}$ is as above and $\bar{b}$ is the firing rate of a local population above background when a single grating is presented to both eyes, maximized over the grating's orientation.} BM, and the simpler ratio $\bar{m}/\bar{b}$, are plotted for all local populations in a HC in Fig.~7.

These plots follow a pattern predicted by our analysis under Finding 2: At \ktext{$\prob(Crs)\geq0.75$}, $\bar{m}/\bar{b}$ exceeds $1$ by $15\%$ on average; it can be as large as \ktext{$30\%$}. At $\prob(Crs)=0$, this ratio is much too small for $20\%$ of the populations\ktext{; see, e.g., \cite{cox2019temporal,dougherty2019binocular}}.  We are thus left with \ktext{$0.25\leq\prob(Crs)\leq0.5$} as the most likely scenario.  Notice that for $\prob(Crs)$ in the \ktext{$0.25$--$0.5$} range, $\bar{m}$ is, on average, $< 10\%$ larger than $\bar{b}$, consistent with what has been reported in, e.g.,~\cite{cox2019temporal,dougherty2019binocular}.  For a neuron located on the border, $\prob(Crs)=1$ means half of its projections are to the opposite column\ktext{, and $0.25\leq\prob(Crs)\leq0.5$} means that about \nicefrac18 to \nicefrac14 of its projections are to the opposite column.

\begin{figure}[tp!]
  \begin{center}
    \ifdraft{}{\includegraphics*[width=\textwidth]{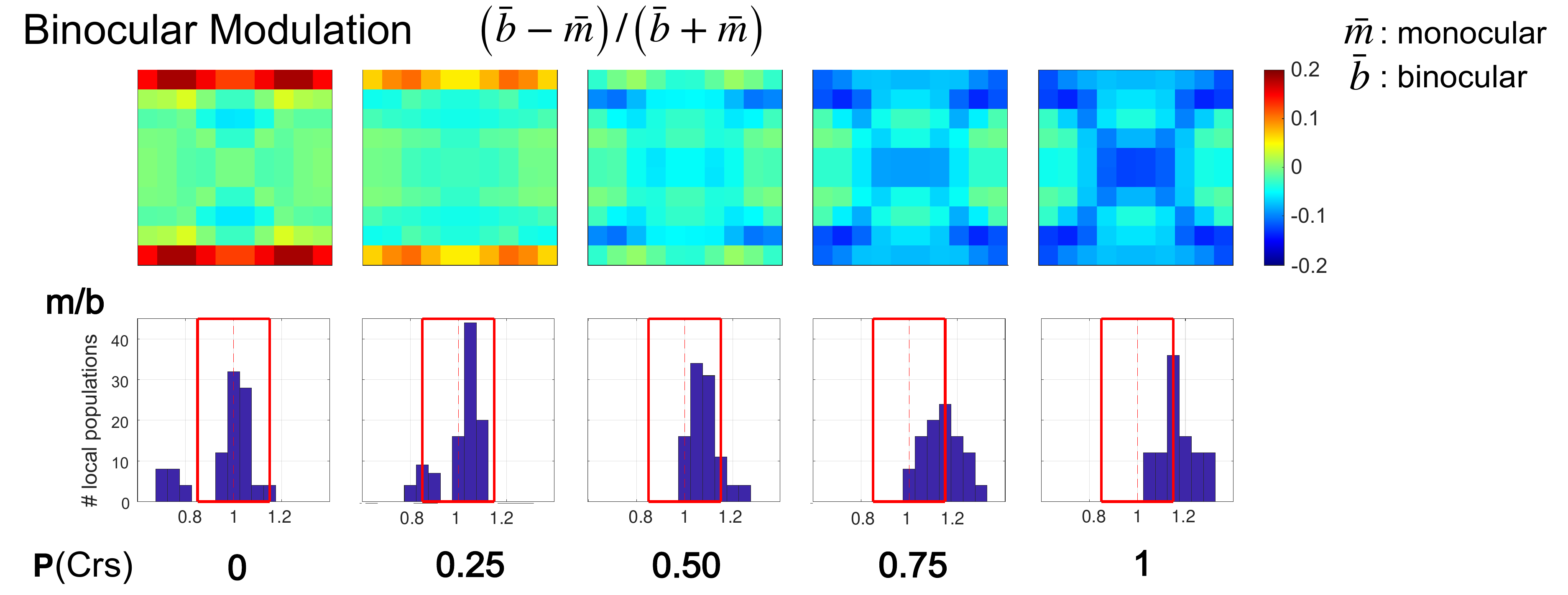}}
    \caption{{\bf Comparison of monocular and binocular responses.} Binocular modulation (top row, see text for definition) and histograms of $\bar{m}/\bar{b}$, the ratio of firing rates when monocularly and binocularly stimulated (bottom row) are shown, with red squares identifying regions where $\bar{b}$ and $\bar{m}$ are within $15\%$ of one another.
    }
     \label{Fig6:Bino-index}
  \end{center}
\end{figure}

\paragraph{Finding 5: Feedback projections from L6 to L4 are largely monocular.}
L6 is one of the more enigmatic layers, with diverse cell types whose functions and dynamics are among the least understood --- even as the anatomy of L6 has by now been well documented \cite{wiser1996contributions,briggs2010organizing,thomson2010neocortical}.  Specifically, L6 is known to have a wide range of cells ranging from almost purely monocular to strongly binocular \cite{wiser1997ocular,dougherty2019binocular,schiller1976quantitative}\knote{Added the Schiller-Finlay-Volman ref from Hawken}, but little information on the binocularity of its {\it projections to L4} is available thus far from experiments. We now use the model to learn about these projections, to better assess how L4 acquires its binocularity.

Up to now, we have assumed in our model that for each stimulus, the magnitudes of L6 projections to L4 are roughly proportional to L4 activity level at corresponding locations. Implicit in this assumption is that L6 projections are as monocular, or binocular, as L4. We now explore the possibility that L6 feedback may be substantially more binocular.  

For a drifting grating and a local population $p$, \ktext{let $m_6(p)$ be the steady-state L6 input to $p$ when its home column is monocularly stimulated, and $b_6(p)$ the steady response under binocular stimulation using the same grating (see SI).}\knote{this ok?}  To simulate a L6 feedback that is ``$x\%$ more binocular than L4,'' we \ktext{set the L6 input into $p$ to}
$$
\frac{x}{100}b_6(p) + \left(1-\frac{x}{100}\right)m_6(p)\ .
$$ 
That is, $x=0$ means L6 feedback is as monocular as L4; $x=100$ means it is purely binocular.

\begin{figure}[tp!]
  \begin{center}
    \begin{subfigure}{\textwidth}
      \ifdraft{}{\includegraphics*[width=\textwidth]{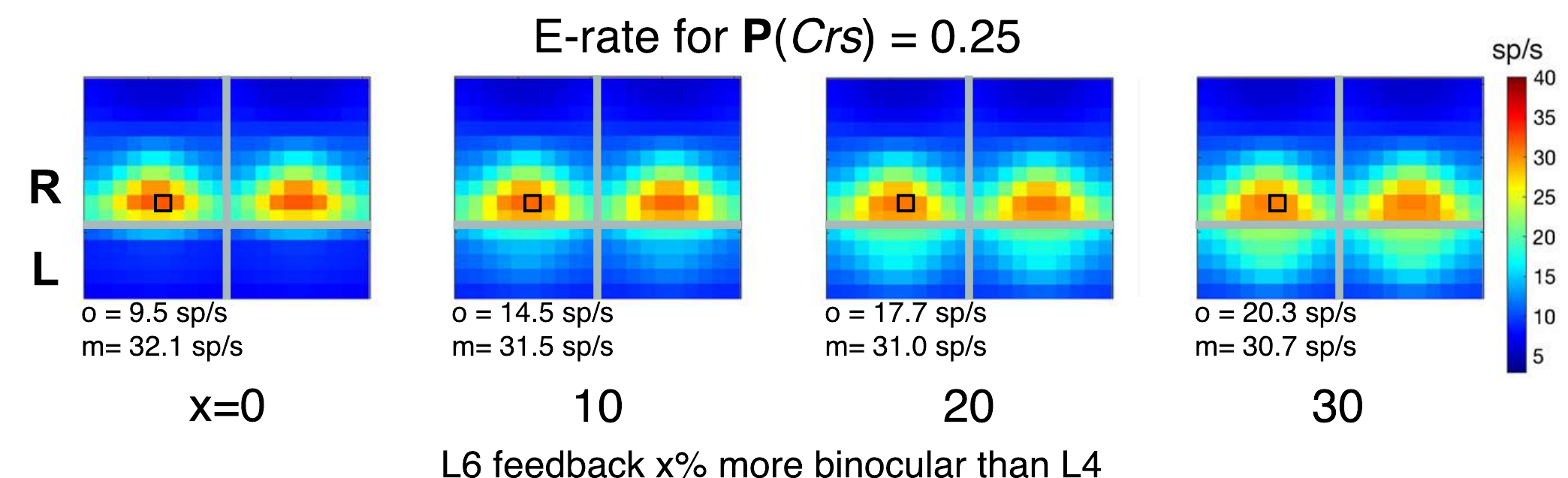}}
    \end{subfigure}
    \caption{{\bf Biologically plausible ranges of L6 feedback binocularity.}~~ Activity maps are shown for various levels of binocularity in L6 feedback to L4.  \ktext{Here the setup is as in Fig.~4, except that the drifting grating presented to the right eye (ODCs in top row) is the 45$^\circ$ grating.}  Firing rates for a local population (small black box) are indicated below the plot: $m$ is the local population's monocular response when its home column is stimulated, $o$ when the other eye is stimulated.}
    \label{Fig8:Bino-index}
  \end{center}
\end{figure}

Fig.~8 shows activity maps under monocular stimulation similar to those in Fig.~4 for a range of $x$ for $\prob(Crs)=0.25$. Only the ODC containing the upper two HCs are stimulated, but as one can see, the bottom ODC shows increasingly substantial responses as $x$ increases.

For concrete comparisons, we selected a local population (\ktext{small} box with black rim in Fig.~8) one row away from the ODC boundary so as not to be in the binocular strip, and computed its firing rates $m$ and $o$ for various values of $x$. The results are shown below each panel.  Notice that $o$ for the selected population is the same as the \ktext{firing rate $m$} of its mirror image \ktext{across the ODC boundary}.
These plots show clearly that BI is more realistic when L6 feedback is $<10\%$ more binocular than L4.

Studies for $\prob(Crs)=0.5$ show similar results; see SI.

\vskip .2in
\paragraph{Summary of binocularity study}  

\medskip
The following results were obtained with the aid of a new multi-scale computational model:

\begin{itemize} 
\item[(i)] While L4 neurons are well known to be largely monocular, the degree of binocularity varies.  In particular, there is a narrow binocular strip along ODC borders, caused largely by cross-ODC projections from L6.
\item[(ii)] Microcircuits in which L4 neurons near ODC borders send \ktext{$10-30\%$} of their projections to the opposite column and the rest to their home column seem most consistent with data.
\item[(iii)] L6 feedback to L4 is largely monocular.
\end{itemize}
Mechanistic explanations are provided for many of the phenomena observed.  The existence of a binocular strip (Item (i)) is an emergent property of the model and is known to be consistent with data.  We propose our other findings as model predictions, to be investigated further in experiments.

\kpart{Discussion}

We have presented a multiscale model of the Macaque V1 aimed at bridging cortical structure and function. Using this model, we have tried to offer theoretical insights, and to produce inferences and testable predictions about unknown anatomy and physiological parameters.

\paragraph{Model-experiment synergy, and relation to previous work}

\medskip \noindent
{\it A comprehensive and scalable model of V1} 

\smallskip
Our model reproduces several hallmark response properties of the input layer (Layer $4C\alpha$) of primate V1, including orientation selectivity, responses to increasing contrast, and the merging of signals from the two eyes in early visual processing \cite{hubel1968receptive,albrecht1982striate,henry2013functional,callaway2005structure,lund2003anatomical}. These results in turn depend on E-I balance, effective interaction within and across ODCs within L4, and the integration of feedforward LGN input and feedback from L6 \cite{douglas1995recurrent,wiser1996contributions}. 

Our work descended from \cite{chariker2016orientation,chariker2020contrast,chariker2022computational}, which together with their predecessors (e.g., \cite{mclaughlin2000neuronal,mclaughlin2003large,tao2004egalitarian}) have substantially advanced computation in primate V1 \cite{potjans2014cell}. The models in this series are more comprehensive and include more biological details than most existing models of V1.  A limitation is that directly simulating large-scale network dynamics is time-consuming, making the models hard to tune and challenging to scale up.

What we have presented in this paper is a new generation of models. Using the multiscale strategy developed in \cite{xiao2024efficient}, we have found a way to evolve a model that is reduced in complexity --- without sacrificing essential biological details --- enabling us to gain substantial speed and scalability.
To demonstrate the fidelity of this modeling approach, we have, in the first part of the paper, reproduced a sample of the results in \cite{chariker2016orientation,chariker2020contrast,chariker2022computational} on key V1 functions. The rest of the paper is then devoted to demonstrating how the model we have built can potentially be used as a discovery tool to advance visual neuroscience.

\medskip \noindent 
{\it Inferred cross-ODC structures and dynamics}

\smallskip
It has been hypothesized that the two streams of signals passed along by the two eyes merge somewhere in V1, but neural substrates for this convergence have yet to be clarified \cite{horton1998monocular}. We explored some simple motifs of neuronal interactions between ODCs inspired by \cite{katz1989local} and others, and found that a surprisingly simple anatomical picture supports the largely but not exclusively monocular responses of neurons seen in Layer $4C\alpha$: of the potential lateral connections, \ktext{$\sim10-30\%$} cross ODC boundaries; the rest are confined to their home columns.  Another question has to do with the feedback from L6. Our model suggests that L6 projections to L4 are mostly monocular; it also points to the experimental fact that these projections can cross ODC borders freely \cite{wiser1997ocular} as being pivotal to the emergence of binocular capabilities of neurons in this input layer.

The findings above exemplify the tight synergy between models and experiments: anatomical insights are incorporated into model design; the model was then used to probe functional consequences, resulting in the generation of hypotheses for cross-column integration, which we present as testable predictions.

\medskip \noindent 
{\it Emergent binocular border strips and experimental validation}

\smallskip
One of the most striking findings from the model is the emergence of narrow binocular strips $\sim100\mu$m in width along the boundaries separating adjacent ODCs, a phenomenon first observed in the laboratory in \cite{horton1998monocular}, which reported the presence of such strips in the Macaque V1.  Indeed these binocular border strips were not evident under normal conditions because both eyes are active; they were unmasked only by perturbations such as monocular enucleation or eyelid suture \cite{horton1998monocular}, which produced alternating patterns of cytochrome oxidase activity revealing the compartments.

We stress that the information in \cite{horton1998monocular} was not used in any way in our model design, and that the appearance of these strips is a purely emergent phenomenon. That our findings align quantitatively with biology (up to tens of microns) can be seen as validation of our modeling \cite{horton1998monocular}.

\paragraph{Broader significance}

\medskip \noindent
{\it Multiscale approaches in neuroscience}

\smallskip

Multiscale methods, which originated in the physical sciences and engineering, provide a suite of mathematical and computational tools for bridging dynamics across vastly different spatial and temporal scales.  A purpose of this paper is to advocate the use of multiscale ideas in neuroscience modeling \ktext{(see \cite{xiao2024efficient} and references therein)}.  In this context, multiscale methods call for representing the system of interest not so much on disparate spatial or temporal scales but on different anatomical scales or levels of biological detail (e.g., cellular, local circuit, laminar, cortical region, $\cdots$), so that the dynamics on each scale can be treated separately, informed by results from other scales \cite{weinan2003heterogeneous,kevrekidis2003equation}. This approach is not always feasible, but when successfully implemented, it can be powerful \cite{abdulle2012heterogeneous,dura2019netpyne,presigny2022colloquium,d2022quest}.

The model used in this paper has two scales: a {\it coarse-grained system} the basic units of which are local populations, and {\it local dynamics}, i.e., dynamics within these populations. The gain in computational speed we mentioned earlier is obtained by leveraging the similarity in \ytext{local} anatomical structures within a single layer of a cortical region to handle their local dynamics more efficiently, eliminating the enormous waste in duplicating similar computations during direct network simulations \cite{xiao2024efficient}. A high-level summary of how this works is given in \textbf{Materials \& Methods}.

As multiscale models allow one to focus on individual scales, we propose that they are especially suited for purposes of inferring information on anatomical structures at specific scales. In this paper we have used observed properties of neurons, namely their responses to monocular and binocular stimulations, to infer the microcircuitry that underlies cross-column interaction.  One can imagine using information on microcircuits to help infer structures and dynamics on larger sales, \ktext{and on up,} leading eventually to systems level understanding \cite{haueis2022descriptive,d2022quest}.

\medskip \noindent
{\it Next steps} 

\smallskip
In neuroscience today, local information tends to be more readily available than relevant facts about long-range interactions \cite{markov2014anatomy}. In V1, for example, local circuits are better understood than long-range connections within layers or interlaminar projections \cite{callaway1998local}, not to mention interactions with higher visual cortical areas. Yet many very basic visual properties of neurons, such as surround suppression, binocularity, direction selectivity, and spatial frequency preferences \cite{skottun1987effects,angelucci2017circuits,wang2020laminar,dougherty2019binocular,cox2019temporal}, are acquired, amplified and/or fine-tuned through nonlocal, multi-stage processing.  Given the capability of the methodology we have in our hands to probe anatomical structures and their functional consequences at multiple scales, these are among the next steps we envision.

\subsection*{Resource availability}

The source code for all computational examples can be found at \href{https://github.com/Texense/NYU-Vision-2Drive.git}{https://github.com/Texense/NYU-Vision-2Drive.git}.

\subsection*{Acknowledgements}

ZCX was supported in part by New York University through the Courant Instructorship and the Shanghai Municipal Education Commission through the Shanghai Oriental Talents Program.  LSY was supported in part by DMS-2350184.  KL was supported in part by the Simons Foundation through grant MP-TSM-00002687.

\subsection*{Author contributions}
Conceptualization, Analysis: LSY, ZCX. Methodology, Writing: LSY, ZCX, KL.  Data Curation and Software: ZCX.

\subsection*{Declaration of interests}
The authors declare no competing interests.

\subsection*{Declaration of generative AI and AI-assisted technologies}

Generative AI technologies were not used in preparing the manuscript or in conducting the work described here.

\subsection*{Supplemental information titles and legends}

\kpart{Materials \& Methods}

\input{methods.tex}

\bibliographystyle{unsrt}
\bibliography{Binocular.bib}

\end{document}

%% file: methods.tex
Here we summarize the main components of the multiscale modeling
strategy introduced in \cite{xiao2024efficient}, to which we refer
interested readers for details.  Model parameters and choices specific
to this paper are described in SI.  The procedure we propose is
considerably more general than how it is used in this paper; it is not
limited to L4, or even to the visual cortex. For concreteness, however,
we will describe the ideas as they pertain to an arbirary layer,
referred to ``Layer X'', of V1.

\newcommand{\thelayer}{Layer X}
\renewcommand{\hat}[1]{\widehat{#1}}

\bigskip \noindent
{\bf Model overview}
\begin{itemize}
\item[1)] {\it Network design.}  First we draw up a blueprint for a
  network model of \thelayer\ incorporating as much of the neuroanatomy
  and physiology as is relevant -- without concerning ourselves at this
  stage with implementation details.  Relevant biological facts
  generally include (i) the specification of cell types within
  \thelayer, their known characteristics, cell densities, connection
  probabilities, etc., and (ii) input sources that originate from
  outside of this layer, e.g., from thalamus, other layers of V1, other
  cortical areas, and nonsynaptic (e.g., cholinergic) inputs.
  
\item[2)] {\it Coarse-graining.}  We then divide \thelayer\ into local
  populations of nearby neurons, referred to as ``pixels,'' basic units
  on which the coarse-grained (CG) model operates.  Neurons of
  each type in a local population are assumed to be identical: their
  dynamics are governed by identical equations, they receive identical
  inputs, and produce identical outputs. Local population sizes should
  be chosen so they are small enough that the assumption above is
  reasonable, and large enough to gain computational efficiency.

\item[3)] {\it A reduced model reflecting biological details.}  Our CG
  model is a rate model the interaction kernels of which are derived
  from neuronal interactions and external inputs as described in Item~1.
  Let $\mathcal P$ be the set of all pixels as defined in Item~2, and
  let $\mathcal J$ be the set of neuron types considered (e.g.,
    simple and complex E-cells, or different types of I-cells).  A
  firing rate configuration is a collection of numbers of the form
  \begin{displaymath}
    f \ = \ \{f_{j,p}, \ j \in \mathcal J, p \in \mathcal P\}\ .
  \end{displaymath}
  When presented with a stimulus, we iteratively update firing
    rate configurations until a steady state is reached.  An outline of
  how firing rate configurations are updated is given
  below.
 \end{itemize} 
 
\noindent
{\bf Updating procedure}
\begin{itemize}
 \item[4)] {\it Interaction kernels.}  Given $p, p' \in \mathcal P$ and
   $j, j' \in \mathcal J$, let $C_{j \leftarrow j'; p \leftarrow p'}$
   denote the number of neurons in local population $p'$ of type $j'$
   presynaptic to neurons of type $j$ in local population $p$ ---
     according to the blueprint in Item~1. Then given $f$, the number
   of spikes $n_{j,p}(j')$ from neurons of type $j'$ received by neurons
   of type $j$ in local population $p$ in the next time step is given by
  \begin{displaymath}
    n_{j,p}(j') = \sum_{p' \in \mathcal P} C_{j \leftarrow j'; p \leftarrow
      p'} \ f_{j',p'}\ .
  \end{displaymath}

\item[5)] {\it External inputs.} Let $\mathcal S$ denote the set of all
  external sources of inputs designated as relevant in Item~1, and let
  $e_{s,j,p}$ denote the quantity of input $s$, $s \in \mathcal S$, to
  be received by neurons of type $j$ in pixel $p$ in the next step.
  These inputs may depend on sensory stimuli or activity elsewhere in
  the brain, and may or may not depend on the previous firing rate
  configuration $f$ of \thelayer.

\item[6)] {\it Updating of firing rate configuration.}  Given $f$ and
  $\{e_{s,j,p}\}$, we now define $\hat f = \{\hat f_{j,p}\}$, the firing
  rate configuration in the next time step.  For definiteness, let us
  assume that individual neurons are modeled as LIF (leaky
  integrate-and-fire) neurons as is done in this paper (though
    other models can be used).  As all neurons of type $j$ in local
  population $p$ are assumed to be identical, we represent them by a
  single LIF neuron. We then define $\hat f_{j,p}$ to be the steady
  state firing rate of this LIF neuron in response to
  \begin{itemize}
  \item[(i)] inputs from \thelayer\ from Item~4 above, i.e., $
    \sum_{j' \in \mathcal J} n_{j,p}(j'),$ and
  \item[(ii)] external inputs as defined in Item~5, i.e., $\sum_{s \in
    \mathcal S} e_{s,j,p}\ .$
  \end{itemize}
  Inputs for which spike times matter may be treated differently; all
  other synaptic inputs are delivered as spikes arriving at Poisson
  times.

\item[7)] {\it Precomputation of potential local responses.} Even though the
  simulation of a single LIF neuron is much faster than simulating the
  entire network, there is a great deal of redundancy in the computation
  of local responses: similar computations are performed for each local
  population at each time step, for each set of initial condition and
  visual stimulus. We proposed in \cite{xiao2024efficient} to precompute
  and tabulate all potential local responses {\it once}, then look up
  and interpolate local responses as needed.
\end{itemize}
{\em Remark on Item~7.}  The actual local responses depend on
circumstance and history; it is not possible to know the response of a
local population to a given stimulus in advance.  What we compute in
Item~7 above is a set of {\it potential} local responses covering all
combinations of firing rates that can occur within reason.
We compute in advance responses to all input values on a grid covering
this feasible region.  These local response functions can be functions
of large numbers of variables (in this paper, one for each cell type in
L4, plus one for each external input source where ``external'' means
outside L4), but as local responses typically vary smoothly and
monotonically with parameters, the grid often need not be too refined.
While this still requires significant resources, it need only be done
once.  The results can be used thereafter for any number of simulations
and, as we have shown, for parameter exploration.